\documentclass[12pt]{article}

\usepackage[dvips]{graphicx}
\usepackage{a4}
\usepackage{amsmath}

\newcommand{\wt}{\widetilde}
\def\Pf{\mathop{\rm Pf}\nolimits}
\def\mod{\mathop{\rm mod}\nolimits}

\def\Tr{\mathop{\rm Tr}\nolimits}

\begin{document}
\begin{titlepage}
    \begin{normalsize}
     \begin{flushright}
                 UT-03-20\\
                 {\tt hep-th/0306239}\\
                 June 2003
     \end{flushright}
    \end{normalsize}
    \begin{Large}
       \vspace{0.6cm}
       \begin{center}
       \bf
          Strings in a PP-wave background\\
          compactified on ${\bf T}^8$ with twisted ${\bf S}^1$
       \end{center}
    \end{Large}
  \vspace{15mm}

\begin{center}
{\large
           Kota Ideguchi
           \footnote{E-mail :
              ideguchi@hep-th.phys.s.u-tokyo.ac.jp}\quad and\quad
           Yosuke Imamura
           \footnote{E-mail :
              imamura@hep-th.phys.s.u-tokyo.ac.jp}

      \vspace{16mm}

              {\it Department of Physics,}\\[0.7ex]
              {\it University of Tokyo,}\\[0.7ex]
              {\it Hongo, Tokyo 113-0033, Japan}
}
      \vspace{1.5cm}
\end{center}
\begin{abstract}
We study a torus-like compactification of type IIB maximally supersymmetric PP-wave background.
As the most general case, we discuss a ${\bf T}^8$ compactification
of all the transverse directions.
A nontrivial structure of the isometry group requires an additional
light-like compactification.
This additional ${\bf S}^1$ fiber is twisted on the ${\bf T}^8$.
We determine the spectrum of closed strings in this twisted torus
background and compute the thermal partition function.
\end{abstract}

\end{titlepage}

\section{Summary and Conclusions}
For some time, string theory has been recognized as a useful tool for investigation of
Yang-Mills theories.
String theories in various backgrounds are expected to be duals of gauge theories
and we have obtained various information about gauge theories by
studying strings.
Recently, PP-waves have attracted great interest because of two facts
that they are duals of certain subsectors of supersymmetric gauge theories\cite{BMN}
and that string theories in PP-wave backgrounds can be exactly solved\cite{metsaev,metsaevtseytlin}.
The simplest, maximally supersymmetric example of PP-waves
is obtained by taking a Penrose limit of
the near horizon geometry of a number of coincident
D3-branes.

In this paper, we discuss a torus-like compactification of
this maximally supersymmetric PP-wave.
Because the isometry of the PP-wave is non-Abelian,
toroidal compactification is possible only for the
directions corresponding to the Cartan part of the isometry group.
Such compactifications have been already considered in the literature.\cite{Michelson,compact,suga2,MMS,para}
Our purpose is to generalize these works to a
torus-like compactification of non-commuting directions.
As a general case, we consider a compactification of the eight transverse
directions.
In addition,
such a compactification requires the light-like direction to be compactified
because the commutator of two transverse translations gives a light-like translation\cite{Michelson}.
Therefore, the internal space we consider is nine-dimensional.
Although uncompactified maximally supersymmetric PP-wave background is
believed to be a dual description of the large R-charge sector of
the ${\cal N}=4$ supersymmetric $SU(N)$ Yang-Mills theory,
it is not clear if our compactified background has anything to do with
Yang-Mills theories.

Following the usual procedure, we obtain the worldsheet Hamiltonian $H$ and
the worldsheet momentum $P$.
(See (\ref{windingP}) and (\ref{windingH}).)
Because the oscillator part is not affected by the compactification,
we mainly focus on the zero-mode part.
In addition to the occupation numbers for oscillators,
string states are labeled by a set of the winding numbers and the Kaluza-Klein momenta.
Because $H$ does not include the transverse momenta,
the transverse momenta label degenerate states
similar to the lowest Landau level states in ${\bf T}^8$
as pointed out in the previous works\cite{RT,compact,MMS,BOPT}.
Due to the non-commutativity of the transverse space,
only half of eight components of the transverse momentum $\vec p$
can be diagonalized simultaneously.
This fact makes it subtle if we are able to diagonalize the worldsheet momentum $P$,
which includes a term $\vec R\cdot\vec p$ where $\vec R$ is a vector
representing the transverse winding of a string.
We show that we can always choose a commutative set of four
components of the momentum so that $\vec R\cdot\vec p$ is
a linear combination of these.
Therefore, we can use the Virasoro constraint in order to
determine the spectrum without any trouble.

We also compute the thermal partition function of strings on the
compactified background.
Although we have to choose four-dimensional commutative subspace
by hand to pick up commuting four components of the Kaluza-Klein momentum,
the final result does not depend on this choice.
The result is quite similar to the ${\bf S}^1$ compactified case given in \cite{suga2},
and the main change is that the sum over one winding number in the result for ${\bf S}^1$ compactification\cite{suga2} is replaced by the
sum over the eight winding numbers.

This paper is organized as follows.
In section 2, we briefly review the light-cone quantization
of strings in the PP-wave background in a rotating coordinate system,
which is convenient for discussions of the compactification.
In section 3, we explain how we compactify the background in detail.
In section 4, we determine the spectrum of
strings in the compactified background and in section 5
we compute the thermal partition function of strings in the background.

\section{The rotating coordinate of the PP-wave}
In this section we briefly review the light-cone quantization of closed strings
in the (uncompactified) maximally supersymmetric PP-wave background in the
rotating coordinate.
The PP-wave metric is given by
\begin{equation}
ds^2=2dX^+dX^-+\sum_{i=1}^8(dX'^idX'^i-\mu^2X'^iX'^i(dX^+)^2).
\label{ppmetric0}
\end{equation}
When we discuss compactifications of the PP-wave,
it is convenient to use the rotating coordinate defined by
\cite{Michelson}
\begin{eqnarray}
X'^{2a-1}&=&X^{2a-1}\cos(\mu X^+)-X^{2a}\sin(\mu X^+),\nonumber\\
X'^{2a}&=&X^{2a-1}\sin(\mu X^+)+X^{2a}\cos(\mu X^+), \quad
(a=1,2,3,4)\label{transform}
\end{eqnarray}
Then, the potential term proportional to $(X'^i)^2$ disappears
and the metric becomes
\begin{equation}
ds^2=2dX^+dX^-+dX^idX^i-\frac{1}{2}F_{ij}X^idX^jdX^+,
\label{ppmetric}
\end{equation}
where $F_{ij}$ is the following skew-diagonal matrix
\begin{equation}
F_{ij}=\left(\begin{array}{cccccccc}
        & -2\mu \\
        2\mu \\
&& \ddots \\
        &&&& -2\mu \\
        &&& 2\mu \\
        \end{array}\right).
\label{Fij}
\end{equation}
Although the light-like coordinate $X^-$ is often reparameterized as $X^-\rightarrow X^-+f(X^i)$
so that some of the space-like isometries become manifest,
we leave $X^-$ unchanged here.

If we take the light-cone gauge
\begin{equation}
X^+=\frac{p_-}{2\pi T}\tau,
\label{LCgauge}
\end{equation}
the bosonic part of the worldsheet Lagrangian is given by
\begin{equation}
L=\int_0^{2\pi}\frac{d\sigma}{2\pi}\left[
p_-\partial_\tau X^-
-\frac{2\pi T}{2}\partial_\alpha X^i\partial^\alpha X^i
+\frac{p_-}{2}F_{ij}X^i\partial_\tau X^j
\right].
\label{rotlag}
\end{equation}
The worldsheet Hamiltonian and the worldsheet momentum are straightforwardly
obtained from this Lagrangian as
\begin{eqnarray}
H&=&2\pi T\int_0^{2\pi}\frac{d\sigma}{2\pi}\Big[
p_-\partial_\tau X^-+\frac{1}{2}(\partial_\tau X^i\partial_\tau X^i+\partial_\sigma X^i\partial_\sigma X^i)\Big],\label{ham0}\\
P&=&-\int\frac{d\sigma}{2\pi}\Big[p_-\partial_\sigma X^-
+2\pi T\partial_\sigma X^i\partial_\tau X^i\Big].
\label{genwsmomentum}
\end{eqnarray}

The mode expansion of $X^i$ is
\begin{eqnarray}
X^{2a-1}+iX^{2a}=\sum_{n=-\infty}^\infty\frac{1}{\sqrt{(2\pi T)\omega_n}}
\big(a_{n,a}e^{-i(\omega_n+\omega)\tau+in\sigma}
    +b_{n,a}^\dagger e^{i(\omega_n-\omega)\tau-in\sigma}\big)\nonumber\\
(a=1,2,3,4),
\label{Z0expand}
\end{eqnarray}
where $\omega$ and $\omega_n$ are defined by
\begin{equation}
\omega=\frac{\mu p_-}{2\pi T},\quad
\omega_n=\sqrt{\omega^2+n^2}.
\end{equation}
The shift of the frequency by $\omega$ is due to the rotation of the
coordinate system.

The action of the fermionic sector is obtained straightforwardly by use of an appropriate 
covariant derivative\cite{metsaevtseytlin}.
The transformation (\ref{transform}) changes the spin connection, and
induces an additional mass term
$-i \frac{\omega}{2} \theta \gamma^{2a-1}\gamma^{2a} \theta$.
This term shifts fermion masses by $\pm\omega/2$.
In the case of a ${\bf T}^8$ compactification, there are contributions from four
rotating planes and
the mass eigenvalues of the eight fermions are
one $\omega_n + 2\omega$, six $\omega_n$'s, and
one $\omega_n-2\omega$.
Although this mass shift breaks the boson-fermion degeneracy,
it does not affect the zero-point energy because
the shift of the zero-point energy is proportional to
the sum of the mass eigenvalues, which is kept invariant
under the coordinate transformation (\ref{transform}).
Since the fermionic sector is not involved by compactifications,
we will ignore the fermionic sector in the rest of this paper.

The bosonic oscillators
satisfy the commutation relations
\begin{equation}
[a_{m,a},a_{n,b}^\dagger]=[b_{m,a},b_{n,b}^\dagger]=\delta_{mn}\delta_{ab},\quad
\mbox{others}=0.
\label{abcomm}
\end{equation}
The Hamiltonian (\ref{ham0}) and the momentum (\ref{genwsmomentum}) are represented by these oscillators as
\begin{eqnarray}
H&=&
\frac{1}{2\pi T}p_-p_+
+\sum_n(\omega_n+\omega)a_{n,a}^\dagger a_{n,a}+\sum_n(\omega_n-\omega)b_{n,a}^\dagger b_{n,a},\\
P&=&\sum_{n,a}n(a^\dagger_{n,a}a_{n,a}+b^\dagger_{n,a}b_{n,a})
\label{ucwsmom}.
\end{eqnarray}
We can now determine the mass spectrum of strings by the Virasoro constraint
$H=P=0$.

As investigated in Refs. \cite{RT,BOPT} for NS-backgrounds and mentioned in \cite{compact} for the RR case, this system is regarded as a system of charged strings moving in the constant gauge flux $F_{ij}$ given by (\ref{Fij}).
To make this fact manifest, we rewrite the Lagrangian (\ref{rotlag}) in the
following form.
\begin{equation}
L=\int_0^{2\pi}\frac{d\sigma}{2\pi}\left[
p_-\partial_\tau X^-
-\frac{2\pi T}{2}\partial_\alpha X^i\partial^\alpha X^i
+p_-(A_i\partial_\tau X^i)
\right].
\label{gaugelag}
\end{equation}
We introduced $A_i$ by
\begin{equation}
A_{2a-1}=\mu X^{2a},\quad
A_{2a}=-\mu X^{2a-1},\quad
(a=1,2,3,4)
\label{bkpot}
\end{equation}
and these are related to $F_{ij}$ by $F_{ij}=\partial_iA_j-\partial_jA_i$.
From this viewpoint, the light-cone momentum $p_-$
is identified with the total charge of a closed string.
The Lagrangian (\ref{gaugelag}) is invariant under the gauge transformation
\begin{equation}
X^-\rightarrow X^--\lambda,\quad
A_i\rightarrow A_i+\frac{\partial}{\partial X^i}\lambda,
\label{gaugetr}
\end{equation}
where $\lambda=\lambda(X^i)$ is a parameter depending on the transverse coordinates $X^i$.

In the expansion (\ref{Z0expand}), we have two $\sigma$-independent modes $a_{0,a}$ and $b_{0,a}$.
These describe motion of the center of mass of a closed string.
The $a_{0,a}$ term has a factor $e^{-2i\omega\tau}$ and
represents the cyclotron motion of the string.
The eigenvalues of the operator $a_{0,a}^\dagger a_{0,a}$ determine
the radius $r_a$ of the cyclotron motion on the $X^{2a-1}$-$X^{2a}$ plane
by the relation
\begin{equation}
r_a^2\sim\frac{1}{(2\pi T)\omega}a_{0,a}^\dagger a_{0,a},
\quad
(a=1,2,3,4).
\end{equation}
On the other hand, we identify $b_{0,a}^\dagger$ term
with a coordinate $x^i$ of the center of the
cyclotron motion by the relation
\begin{equation}
x^{2a-1}+ix^{2a}=\frac{1}{\sqrt{(2\pi T)\omega}}b_{0,a}^\dagger,
\quad
(a=1,2,3,4).
\label{eq18}
\end{equation}
From the commutation relation (\ref{abcomm}), the $x^i$-space possesses the following non-commutativity.
\begin{equation}
i[x^i,x^j]
=\frac{1}{p_-}(F^{-1})^{ij}.
\label{xycommute}
\end{equation}
Because of this commutation relation,
we can represent the momentum $p_i$ conjugate to $x^i$
by the coordinate $x^i$ itself as
\begin{equation}
p_i=p_-F_{ij}x^j.
\label{pandx}
\end{equation}

\section{Compactification}
Let us discuss a ${\bf T}^8$ compactification of the transverse space ${\bf R}^8$.
Let $\vec b_A=(b_A^1,\ldots b_A^8)$ ($A=1,\ldots,8$) form a basis of the compactification lattice
on ${\bf R}^8$.
In many works studying compactifications of PP-waves, each vector $\vec b_A$ is taken in one of $X^{2a-1}$-$X^{2a}$ planes.
Now, however, $\{\vec b_A\}$ can be an arbitrary basis
in ${\bf R}^8$ except that they must satisfy the following
flux quantization condition.
\begin{equation}
p_- (b_A^i F_{ij} b_B^j)\in2\pi{\bf Z}.
\label{Fquant}
\end{equation}
This is necessary for the string wave function on the torus to be well-defined.
If the dimension of the basis was equal to or less than $4$,
we could take a basis satisfying $b_A^i F_{ij} b_B^j=0$.
Such a basis gives a commutative toroidal compactification,
which has been already studied in the literature\cite{Michelson,MMS}.
In the case of the ${\bf T}^8$ compactification, however,
it is impossible to take such a basis, and
we necessarily have to assume the quantization
of the charge $p_-$.
This implies that the light-like direction $X^-$ have to be compactified.
Once $X^-$ direction is compactified with radius $R^-$,
the charge $p_-$ is quantized as
\begin{equation}
p_-=\frac{k}{R^-},\quad
k\in{\bf Z},
\label{llquantized}
\end{equation}
and the
condition (\ref{Fquant}) is rewritten as
\begin{equation}
\Phi_{AB}\equiv\frac{1}{2\pi R^-}(b_A^iF_{ij}b_B^j)\in{\bf Z}.
\label{Fquant2}
\end{equation}

The necessity for the light-like compactification and the flux
quantization condition (\ref{Fquant2}) is also explained
from the structure of the isometry group of the PP-wave\cite{Michelson}.
Let $P_i$ and $P_-$ denote shift operators along $x^i$ and $x^-$, respectively.
These are related to the notation in \cite{Michelson} as
\begin{equation}
P_{2a-1}=-k_{s^+_{2a-1,2a}},\quad
P_{2a}=-k_{s^-_{2a,2a-1}},\quad(a=1,2,3,4),\quad
P_-=-k_{e_-}.
\end{equation}
In the rotating coordinate, these isometries are represented as
\begin{equation}
P_i=\partial_i+\frac{1}{2}F_{ij}X^j\partial_-,\quad
P_-=\partial_-,
\end{equation}
and satisfy the commutation relation
\begin{equation}
[P_i,P_j]=-F_{ij}P_-.
\label{ppcomm}
\end{equation}
Because of the non-vanishing commutation relation (\ref{ppcomm}),
the commutator of two space-like translations
$\exp(b_A^iP_i)$ and $\exp(b_B^iP_i)$
gives the light-like translation:
\begin{equation}
e^{b_A^iP_i}
e^{b_B^iP_i}
e^{-b_A^iP_i}
e^{-b_B^iP_i}
=\exp(-b_A^iF_{ij}b_B^jP_-).
\end{equation}
This implies that the light-like direction $X^-$ must be compactified
by a certain radius $R^-$ and $b_A^iF_{ij}b_B^j$ must be a
multiple of the period $2\pi R^-$ for any pair of $\vec b_A$ and $\vec b_B$.
This reproduces the constraint (\ref{Fquant2}).

Let $G$ be an orbifold group generated by elements $\exp(b_A^iP_i)$
and $\exp(2\pi R^-P_-)$. An arbitrary element of $G$ is represented as
\begin{equation}
g(w^A,n)=e^{w^Ab_A^iP_i}e^{2\pi(n+\sigma(w^A)/2)R^-P_-},
\end{equation}
where $\sigma(w^A)$ is an integral function
satisfying
\begin{equation}
\sigma(w^A+w'^A)=\sigma(w^A)+\sigma(w'^A)-w^A\Phi_{AB}w'^B\quad\mod 2.
\end{equation}
For example, we can define $\sigma(w^A)$ by
\begin{equation}
\sigma(w^A)=\sum_{A<B}w^A\Phi_{AB}w^B.
\end{equation}
The multiplication rule is
$g(w_1^A,n_1)g(w_2^A,n_2)=g(w_3^A,n_3)$
with $w_3^A$ and $n_3$ defined by
\begin{eqnarray}
&&w_3^A=w_1^A+w_2^A,\nonumber\\
&&n_3=n_1+n_2+\frac{1}{2}(-w_1^A\Phi_{AB}w_2^B+\sigma(w_1^A)+\sigma(w_2^A)-\sigma(w_3^A)).
\label{multiplication}
\end{eqnarray}

In general,
compactifications break supersymmetries which are changed by
compactification isometries. On a rotating plane $X^{2a-1}$-$X^{2a}$,
isometries used by the
compactification are $k_{S^+_{2a-1,2a}}$ and $k_{S^-_{2a,2a-1}}$. 
Both the isometries
preserve the same 24 supersymmetries; for instance $k_{S^+_{12}}$ or
$k_{S^-_{21}}$ preserves supersymmetries corresponding to the killing spinors
which vanish by the action of $\gamma^+ ( 1+ i\gamma^3\gamma^4)$\cite{Michelson}. So, when we
choose general directions
of the compactification, the number of preserved supersymmetries depends only
on the number of rotating planes which are
involved by the compactification (See Table \ref{susy}.).
\begin{table}[htb]
\begin{center}
\begin{tabular}{ccc}
\hline
\hline
\shortstack{The number of planes\\involved by compactification}&
 \shortstack{The number of preserved\\supersymmetries}&
 \shortstack{Possible maximal\\compactification}\\
 \hline
1 & 24 & $T^2$\\
2 & 20 &$T^4$\\
3 & 18 &$T^6$\\
4 & 16 &$T^8$\\
\hline
\end{tabular}
\end{center}
\caption{The number of supersymmetries preserved by compactifications}
\label{susy}
\end{table}

\section{Quantization of winding sectors}
Let us consider a winding sector with the boundary condition
\begin{equation}
X^\mu(\sigma+2\pi)=g(w^A,n)X^\mu(\sigma)g^{-1}(w^A,n),
\label{gXg}
\end{equation}
where $w^A$ and $n$ are the transverse and light-like winding numbers,
respectively.
For the space-like and light-like coordinates, (\ref{gXg}) represents the
following boundary conditions, respectively.
\begin{eqnarray}
X^i(\sigma+2\pi)&=&X^i(\sigma)+2\pi R^i,
\label{Z'bc}\\
X^-(\sigma+2\pi)&=&X^-(\sigma)+2\pi n R^-+\pi F_{ij}R^iX^j,
\label{X-bc}\\
X^+(\sigma+2\pi)&=&X^+(\sigma),
\end{eqnarray}
where we defined $R^i=w^Ab_A^i/2\pi$.

First we have to represent the Hamiltonian $H$ and the momentum $P$
by the creation and annihilation operators, the zero-mode momenta and
the winding numbers.
This is straightforward and
almost parallel to the uncompactified case.
We just mention several points we should be careful of.

We decompose $X^i(\sigma)$ into the periodic part $X_0^i$ and the winding part $R^i\sigma$ by
\begin{equation}
X^i(\sigma)=X_0^i+R^i\sigma,
\label{zz0r}
\end{equation}
and the periodic part is expanded by (\ref{Z0expand}).
We can obtain $H$ and $P$ in terms of
the oscillators, the winding numbers and the momenta by substituting the expansion
into (\ref{ham0}) and (\ref{genwsmomentum}).
Then, we should bear in mind that some quantities are not periodic in
$\sigma$.
For example, substituting (\ref{zz0r}) into the worldsheet momentum (\ref{genwsmomentum}), we obtain
\begin{eqnarray}
P
&=&\int_{\sigma_0}^{\sigma_0+2\pi}\frac{d\sigma}{2\pi}\Big[-p_-\partial_\sigma X^-
-2\pi T\partial_\sigma X_0^i\partial_\tau X_0^i
-\frac{p_-}{2}F_{ij}X_0^i\partial_\sigma X_0^j
\nonumber\\&&
-2\pi TR^i\partial_\tau X_0
+\frac{p_-}{2}F_{ij}R^i X_0^j
-\frac{p_-}{2}\sigma F_{ij}R^i\partial_\sigma X^j
\Big].
\end{eqnarray}
Because $X^-$ is not periodic,
we cannot drop the first term.
Similarly, the last term in this expression is not periodic.
In fact, the non-periodicity of these two terms cancels out each other.
To see this, we integrate the last term by part, and obtain
\begin{eqnarray}
P
&=&\int_{\sigma_0}^{\sigma_0+2\pi}\frac{d\sigma}{2\pi}\Big(
-2\pi T\partial_\sigma X_0^i\partial_\tau X_0^i
-\frac{p_-}{2}F_{ij}X_0^i\partial_\sigma X_0^j
\nonumber\\&&
-2\pi TR^i\partial_\tau X_0^i
+p_-F_{ij}R^iX_0^j
\Big)d\sigma
\nonumber\\&&
-\frac{p_-}{2\pi}\Big[X^-+\frac{1}{2}\sigma F_{ij}R^iX^j\Big]^{\sigma_0+2\pi}_{\sigma_0}.\label{eq38}
\end{eqnarray}
The first line in (\ref{eq38}) does not depend on $R^i$ and is equal to the
worldsheet momentum for the uncompactified PP-wave (\ref{ucwsmom}).
The second line includes only the time independent part ($b_0$-term) of $X^i$,
and represents the winding number contribution to the level matching condition.
The third line comes from the integration by part.
Although each term in the bracket in the third line depends on a choice of
the interval $[\sigma_0,\sigma_0+2\pi]$,
the sum of them is independent of $\sigma_0$
due to the boundary conditions (\ref{Z'bc}) and (\ref{X-bc}).
As a result, we obtain
\begin{equation}
P=-p_-R^-+\sum_{n,a}n(a_{n,a}^\dagger a_{n,a}+b_{n,a}^\dagger b_{n,a})-p_iR^i.
\label{windingP}
\end{equation}
We have used the equivalence (\ref{pandx}) between the coordinates and the momenta
to obtain the last term in (\ref{windingP}).

The worldsheet Hamiltonian is obtained in a similar way as
\begin{equation}
H
=\frac{p_+p_-}{2\pi T}+\sum_{n,a}(\omega_n+\omega)a_{n,a}^\dagger a_{n,a}+\sum_{n,a}(\omega_n-\omega)b_{n,a}^\dagger b_{n,a}+\frac{2\pi T}{2}R^iR^i.
\label{windingH}
\end{equation}
In deriving this Hamiltonian, we defined the momentum $p_+$ as a constant part
of $\Pi_+$,
where $\Pi_+$ is the canonical momentum for the field $X^+$:
\begin{equation}
\Pi_+
=\partial_\tau X^-+\frac{p_-}{2(2\pi T)}F_{ij}X^i\partial_\tau X^j.
\end{equation}
Because the periodicity of $\Pi_+$ is guaranteed by the boundary conditions (\ref{Z'bc}) and (\ref{X-bc}),
we can define $p_+$ unambiguously.

We can make an arbitrary string state by acting the bosonic creation oscillators
$a_{n,a}^\dagger$ ($n\in{\bf Z}$), $b_{n,a}^\dagger$ ($n\in{\bf Z}-\{0\}$)
and the fermionic ones on a ground state
\begin{equation}
|n,w^A,k,\vec p_V\rangle,
\label{groundstate}
\end{equation}
in each winding and momentum sector.
$n$, $w^A$ and $k$ are the light-like winding number, the transverse 
winding numbers and the quantum number of the light-like Kaluza-Klein 
momentum(\ref{llquantized}), respectively.
The vector $\vec p_V$ is the transverse momentum in ${\bf T}^8$.

Because the fermionic oscillators
and the bosonic ones except $b_{0,a}$
are not affected by the compactification,
we discuss only the structure of Fock space
associated with the zero-mode oscillator $b_{0,a}$,
which is related to $\vec p_V$ via (\ref{eq18}) and (\ref{pandx}).

The transverse space is non-commutative
and the eight components of $p_i$ do not commute:
\begin{equation}
[p_i,p_j]=-ip_-F_{ij}.
\label{ppcommute}
\end{equation}
Thus we can diagonalize only four of the eight components of
the transverse momentum.
To choose four linearly independent momenta commutative among them,
we define a four-dimensional
commutative subspace $V$ of the transverse space ${\bf R}^8$, and
we use four components of a momentum $\vec p_V$ on $V$ as quantum numbers of ground states.
We assume the subspace $V$ satisfies the following conditions.
\begin{itemize}
\item 
The sublattice $\Gamma_V\equiv V\cap\Gamma_8$
of the ${\bf T}^8$ compactification lattice $\Gamma_8$ is four-dimensional.
This is equivalent to the statement that we can take a basis of $V$
in the lattice $\Gamma_8$.
\item 
Any two momenta along $V$ commute with each other.
Namely, arbitrary two vectors $\vec v_1\in V$ and $\vec v_2\in V$ satisfy
\begin{equation}
\vec v_1\cdot J\cdot \vec v_2=0,
\label{BFB0}
\end{equation}
where $J$ is a skew-diagonal matrix proportional to $F$ and
satisfying $J^2=-1$.
Thanks to this condition,
we can use the four components of $\vec{p}_V$ as independent quantum
numbers specifying ground states.
\item 
The sublattice $\Gamma_V$ includes the winding vector $w^A\vec b_A$.
This guarantees that the term $R^ip_i$
appearing in the worldsheet momentum (\ref{windingP})
is automatically diagonalized on the ground states (\ref{groundstate}).
This condition implies that the subspace $V$ has to be chosen
after the winding numbers $w^A$ are specified.
\end{itemize}

We can always take such a subspace $V$.
Indeed, we can construct a basis of $V$ in the following way.
First, we adopt the winding vector $2\pi\vec R=w^A\vec b_A$
as the first vector $\vec v_1$ in the basis.
If we assume that we have already determined $k$ vectors $\vec v_i\in\Gamma_8$ ($1\leq k<4$) in the basis,
we can choose one solution of the commutativity condition
\begin{equation}
\vec v_{k+1}\cdot F\cdot\vec v_i=0,\quad \mbox{for $i=1,\ldots,k$}
\label{newv}
\end{equation}
as the $k+1$-th vector in the basis of $V$.
Thanks to the flux quantization condition (\ref{Fquant}), all the linearly independent
$8-k$ solutions can be taken in the lattice $\Gamma_8$.
Because this equation has $8-k$ linearly independent solutions,
 we can choose $\vec v_{k+1}$
which is independent of $k$ vectors $\{\vec v_1,\cdots,\vec v_k\}$,
whenever $k$ is less than $4$.
This procedure can be repeated before $k$ reaches $4$,
in which case all the solutions of (\ref{newv}) 
are linear combinations of $\{\vec v_1,\cdots, \vec v_4\}$.
In this way, we obtain four vectors spanning four-dimensional sublattice
of $\Gamma_8$, and we obtain $V$ as a four-dimensional subspace
of ${\bf R}^8$ containing the sublattice.
The above procedure does not guarantee that $\vec v_i$ are
the basis of $V\cap\Gamma_8$.
The lattice spanned by $\vec v_i$ may be a sublattice of $V\cap\Gamma_8$.
However, once we have obtained $V$, we can always choose $\vec v_i$ as a
basis of $V \cap\Gamma_8$.
We will assume $\vec v_i$ to be defined in this way below.

The complement subspace of $V$ in ${\bf R}^8$ is denoted by $W$.
It is given by
\begin{equation}
W=\{\vec w|\vec w=J\vec v,\vec v\in V\}.
\end{equation}
Indeed, $\vec w\in W$ and $\vec v\in V$ are always orthogonal because
$\vec w$ can be represented as $J\vec v'$ for some vector $\vec v'\in V$,
and the inner product $\vec v\cdot\vec w=\vec vJ\vec v'$ is $0$ due to the
assumption (\ref{BFB0}).
In addition, the subspace $W$ is commutative as well as $V$
because arbitrary pair of vectors $\vec w_1=J\vec v_1$ and $\vec w_2=J\vec v_2$
in $W$ satisfies $\vec w_1J\vec w_2=\vec v_1J\vec v_2=0$.

Now, we have obtained a set of quantum numbers specifying string states.
To obtain physical string states, we need to pick up
states invariant under the orbifold group $G$.
This is realized by the identification $g(s^A,t)|s\rangle\sim |s\rangle$
for arbitrary elements $g(s^A,t)\in G$.
The invariance under elements in the form $g(0,t)$
requires the quantization of the light-like momentum.
This is already taken into account by (\ref{llquantized}).
The invariance under
\begin{equation}
g(s^A,0)=\exp(\vec s\cdot\vec P),\quad
\vec s\equiv s^A\vec b_A,
\label{gs0}
\end{equation}
should also be considered.
If $\vec s$ is an element of $V$,
$\vec P$ in (\ref{gs0}) can be replaced by a c-number $i \vec p_V$
when it acts on (\ref{groundstate}), and we obtain
\begin{equation}
g(s^A,0)|n,w^A,k,\vec p_V\rangle
=e^{i \vec s\cdot\vec p_V}|n,w^A,k,\vec p_V\rangle,\quad
\vec s\in\Gamma_V.
\end{equation}
This should be identified with the state (\ref{groundstate})
and it requires the Kaluza-Klein momentum $\vec p_V$ to be quantized
as
\begin{equation}
\vec p_V\in2\pi\Gamma_V^{-1}.
\label{kkquant}
\end{equation}
For the case of $\vec s\notin V$,
it is convenient to decompose $\vec s$
into the sum of two vectors $\vec s_V\in V$ and $\vec s_W\in W$.
If we define $\Gamma_W$ as a projection of $\Gamma_8$ to $W$,
$\vec s_W$ is an element of $\Gamma_W$.
Due to the commutation relation (\ref{ppcommute}),
$\exp(\vec s_W\cdot\vec P)$ shifts the momentum $\vec p_V$ by
$p_-F\cdot\vec s_W$ while $\exp(\vec s_V\cdot \vec P)$
just gives a phase factor when
it acts on the ground states.
For the shift of the transverse vector $\vec p_V$
and the quantization (\ref{kkquant}) of $\vec p_V$ to be consistent to each other,
the vector $p_-F\cdot\vec s_W$ must be an element of the
lattice $2\pi\Gamma_V^{-1}$ as well as $\vec p_V$.
This is actually guaranteed by the flux quantization condition (\ref{Fquant2}).

The action of the orbifold group $G$ also changes the light-like winding number.
This is because the shift replaces the $X$ in the boundary condition
(\ref{gXg}) by $g(s^A,t)Xg^{-1}(s^A,t)$ and it gives new boundary condition
(\ref{gXg}) with $g(w^A,n)$ replaced by $g(w^A,n')$ defined by
\begin{equation}
g^{-1}(s^A,t)
g(w^A,n)
g(s^A,t)
=g(w^A,n').
\label{conjrel}
\end{equation}
The explicit form of $n'$ can be determined by the multiplication rule (\ref{multiplication}) as
\begin{equation}
n'=n+s^A\Phi_{AB}w^B.
\end{equation}
Taking account of these facts, $g(s^A,0)$ changes
the ground state $|n,w^A,k,\vec p_V\rangle$ to
\begin{equation}
e^{\vec s\cdot\vec P}|n,w^A,k,\vec p_V\rangle
=(\mbox{phase factor})\times|n',w^A,k,\vec p_V+p_-F\cdot\vec s_W\rangle.
\label{Wshift}
\end{equation}

The change of the winding number by the action of the orbifold group is a
general feature of non-Abelian orbifolds.
Topologically inequivalent sectors of a non-Abelian orbifold
are labeled not by elements of an orbifold group
but by its conjugacy classes.
In our case, there are the following conjugacy classes.
\begin{eqnarray}
&&\wt g(0,n)=\{g(0,n)\},\quad(n\in{\bf Z}),\nonumber\\
&&\wt g(w^A,n)=\{g(w^A,n+p(w^A)l)|l\in{\bf Z}\},\quad(n=0,1,\ldots,p(w^A)-1),
\end{eqnarray}
where $p(w^A)$ is an integer defined by
\begin{equation}
p(w^A)=\gcd(\Phi_{1A}w^A,\Phi_{2A}w^A,\ldots,\Phi_{8A}w^A).
\end{equation}

Although the topologically inequivalent sectors
are labeled by conjugacy classes $\wt g(w^A,n)$,
we use elements $g(w^A,n)$ of the orbifold group $G$
to label sectors and
treat these sectors independently.
Instead, we restrict the transverse momentum $\vec p_V$ inside a fundamental
region of the lattice $p_- F \Gamma_W$
to take the identification (\ref{Wshift}) into account.
In other words, we treat $\vec p_V$ as a vector in the following compactified lattice.
\begin{equation}
\vec p_V\in 2\pi\Gamma_V^{-1}/(p_-F\Gamma_W).
\label{Fquotient}
\end{equation}

Now we have a complete set of quantum numbers and the restriction on them
to avoid multiple counting of states.
We can obtain the string spectrum by the Virasoro constraint $H=P=0$.

As a consistency check, let us confirm the spectrum
reduces to that of ordinary toroidal compactification
in the vanishing flux limit $\mu \rightarrow 0$.
For the limit to be taken smoothly,
we assume large transverse compactification radii and small
light-like one.
In this situation, the components of the field strength $F_{ij}$ satisfying
the quantization condition (\ref{Fquant2}) can be
treated as continuous quantities.
To show the coincidence of the spectrum,
let us compute the light-cone partition function
$Z_{\rm lc}$ for fixed light-like winding $n$ and light-like momentum $k$ defined by
\begin{equation}
Z_{\rm lc}=\Tr e^{2\pi(-\tau_2H+i\tau_1P)},
\label{Zlcdef}
\end{equation}
where the trace is taken over states with fixed $k$ and $n$.
Because of the large transverse compactification radii,
we can focus only on the $w^A=0$ sector.
For the $w^A=0$ sector, the partition function is factorized into
\begin{equation}
Z_{\rm lc}
=Z_0
{\cal N}_{LLL}\prod_n Z_{n,+}^4\prod_{n\neq0} Z_{n,-}^4,
\label{Zlcfact}
\end{equation}
where $Z_0=e^{2\pi(-\tau_2 p_+p_-/(2\pi T)-i\tau_1 p_-R^-)}$ represents the factor depending on the light-like components of momentum and the light-like winding number.
$Z_{n,+}$ and $Z_{n,-}$ are the factors
coming from the sum over the occupation numbers for the
oscillators $a_{n,a}$ and $b_{n,a}$, respectively,
and are defined as
\begin{equation}
Z_{n,\pm}
=\sum_{N=0}^\infty e^{2\pi(-\tau_2(\omega_n\pm\omega)+i\tau_1n)N}.
\end{equation}
${\cal N}_{LLL}$ is the number of states in the lowest Landau level
labeled by $\vec p_V$.
It is obtained as the ratio between volumes of
the fundamental regions of lattices in the numerator and the denominator
in (\ref{Fquotient}).
\begin{equation}
{\cal N}_{LLL}=\frac{(2p_-\mu)^4{\rm vol}[\Gamma_W]}{(2\pi)^4({\rm vol}[\Gamma_V])^{-1}}
=\frac{(2p_-\mu)^4}{(2\pi)^4}{\rm vol}[\Gamma_8]
=k^4\Pf(\Phi_{AB}).
\label{degeneracy}
\end{equation}
${\rm vol}[\Gamma]$ represents the volume of a fundamental region of
a lattice $\Gamma$.
The final expression shows that this quantity is obviously integer.

On the other hand,
in the case of ordinary toroidal compactification,
the light-cone partition function defied by (\ref{Zlcdef}) is factorized into
\begin{equation}
Z_{\rm lc}
=Z_0
\sum_pe^{ -\tau_2\frac{1}{2T}p^2}
\prod_{n\neq0}Z_{n,0}^8
\label{Zord},
\end{equation}
where $Z_0$ is the same with that in (\ref{Zlcfact}),
and $Z_{n,0}$ is the factor coming from the bosonic oscillators and is given by
\begin{equation}
Z_{n,0}=\sum_{N=0}^\infty e^{2\pi(-\tau_2|n|+\tau_1 n)N}.
\end{equation}
Thanks to the large transverse compactification, the summation over the transverse momentum in (\ref{Zord}) is replaced by an integration as
\begin{equation}
\sum_pe^{ -\tau_2\frac{1}{4\pi T}p^2}
=\frac{{\rm vol}[\Gamma_8]}{(2\pi)^8}\int d^8p
e^{ -\tau_2\frac{1}{2T}p^2}
={\rm vol}[\Gamma_8]\frac{T^4}{(2\pi\tau_2)^4}.
\end{equation}
In the $\mu\rightarrow 0$ limit, both $Z_{n,+}$ and $Z_{n,-}$ reduce into $Z_{n,0}$.
Therefore, for the two partition functions (\ref{Zlcfact}) and (\ref{Zord})
to coincide in the limit, we need the relation
\begin{equation}
{\cal N}_{LLL}Z_{0,+}\stackrel{\mu\rightarrow0}{\rightarrow}
\sum_pe^{ -\tau_2\frac{1}{2T}p^2}.
\end{equation}
Indeed, this relation can be easily shown to hold.

\section{Thermal Partition Function}
In this section, we calculate the thermal partition function (TPF) of strings on our
background.
The computation of this quantity is well done for DLCQ strings. It is
considered that TPF has a property of the modular invariance.
Although this fact is made clear by the path integral calculation of TPF,
we will calculate it by the operator method, because these two methods
are considered to give the same results\cite{GrignaniSemenoff,Semenoff,suga1}. Our
procedure is almost parallel to that of \cite{suga2}, in which ${\bf S}^1$
compactification is treated.

The TPF is defined by 
\begin{eqnarray}
 F(\beta) = 
  -\sum_{l=1}^\infty \frac{1}{\beta l} \textrm{Tr}
  \left[ (-1)^{(l+1) \textbf{F}} e^{-\beta l p^0}\right],
  \label{thermalPF}
\end{eqnarray}
where $\beta$ is the inverse temperature, $\textbf{F}$ is the space-time fermion
number and $p^0 = \frac{1}{\sqrt{2}} (p^+ - p^-)$. The trace runs only
physical states satisfying the level matching condition. 

In the case of an ${\bf S}^1$ compactification, TPF is calculated in \cite{suga2}. The result
is
\begin{align}
F_{{\bf S}^1}(\beta)
 = -\sum_{k=1}^\infty \sum_{l=1}^\infty \sum_{q=0}^{k-1} 
\sum_{w \in (k/d) \textbf{Z}} \frac{{\cal N}}{\beta l k} 
e^{-\frac{\beta ^2 l^2}{4\pi \alpha' \tau_2} 
\left( 1+ \frac{{R^-}^2 R_T^2 w^2}{2 k^2 {\alpha'}^2}\right) }\nonumber\\
\times \textrm{tr}\left[ (-1)^{(l+1)\textbf{F}} e^{-2\pi \tau_2 H_\textrm{osc} 
+ 2\pi i \tau_1 P_\textrm{osc}}\right],\label{S1TPF}
\end{align}
where $R_T$ is the transverse compactification radius,
and $\cal N$ represents the number of the lowest Landau level states on
non-commutative plane compactified on ${\bf S}^1$.
In \cite{suga2}, a cut-off length $L$ for the direction transverse to the
${\bf S}^1$ is introduced to make $\cal N$ finite.
In our case, as we will see below, $\cal N$ is replaced by ${\cal
N}_{LLL}$
in (\ref{degeneracy}),
the number of the lowest Landau level states in ${\bf T}^8$,
which is finite due to the compactness of the ${\bf T}^8$. 
The trace is
over the oscillator modes and $d$ is $\textrm{gcd}(k,q)$. $\tau$ is defined
by 
\begin{equation}
\tau =\tau_1 + i\tau_2 \equiv \frac{q+ il \nu}{k},\quad 
\nu = \frac{\sqrt{2} \beta R^-}{4\pi \alpha'}.\label{tauDef}
\end{equation}
If $k$ is very large, the sum over $l$ and $q$
can be regarded as an integration over
the complex variable $\tau$.

We choose a commutative space $V$ as to contain the transverse winding
vector $w^A\vec b_A$. We can change the basis to $\{\vec{\tilde{b}}_A\}$
whose first four are the four basis vectors of $\Gamma_V$. We denote these four
vectors as $\vec v_i (i=1,2,3,4)$. The other four vectors in $\{
\vec{\tilde{b}}_A\}$ are denoted by $\vec w_i$. 
In this basis, $\Phi_{AB}$ is written in the following form.
\begin{equation}
 \Phi_{AB} = \frac{1}{2\pi R^-} (\vec{\tilde{b}}_A \cdot F \cdot \vec{\tilde{b}}_B)
 = \begin{pmatrix}
    0 & B\cr
    -B & C\cr
\end{pmatrix}.
\end{equation}
It is always possible by basis transformation of $\{\vec{v}_i\},
\{\vec{w}_i\}$ in each subspace to make $B$ diagonal\footnote{Basis
change $\vec{v}$'s and $\vec{w}$'s is each represented by unimodular
matrices. It is known that any integer valued matrix is made integer
valued diagonal matrix by multiplication of two unimodular matrices on
both sides. }. 
In this choice of basis, the matrix $B$ becomes 
\begin{equation}
 B = \begin{pmatrix}
      N_1 & & \cr
       & \ddots & \cr
       & & N_4
\end{pmatrix}.
\end{equation}
The winding vector is represented as 
a linear combination of $\vec{v}_i$'s and we denote the coefficients by $u^i$:
\begin{equation}
\sum_{A=1}^8w^A\vec b_A=
\sum_{i=1}^4u^i\vec v_i.
\end{equation}
We use $R_{\{w^A\}}$ to represent the length of $\sum_{A=1}^8 w^A
\vec{b}_A/2\pi$. 

Because of the projection by the orbifold group,
the ground states are identified as
\begin{equation}
|n,k,w_A,m_i\rangle\sim
|n+\sum_{i=1}^4\alpha_iu_iN_i,k,w^A,m_i+\alpha_ikN_i\rangle,
\end{equation}
where $\alpha_i$ are arbitrary integers.
To avoid multiple counting of states, we restrict $m_i$ by
$0\leq m_i<kN_i$.

In this setting, the physical states are labeled
by the zero-mode quantum numbers in Table \ref{zero-qn}.
\begin{table}[htb]
\begin{center}
\begin{tabular}{ccc}
\hline
\hline
quantum numbers & variables & values\\
\hline
light-cone winding number:& $n$ & $\textbf{Z}$\\
light-cone momentum:& $k$ & $\textbf{N}$\\
transverse winding number:& $w^A$ & $\textbf{Z}$\\
transverse momentum:& $m_i$ $(i=1,\dots,4)$& $\textbf{Z}, 0\leq m_i < kN_i$\\
\hline
\end{tabular}
\end{center}
\caption{The zero-mode quantum numbers}
\label{zero-qn}
\end{table}
The worldsheet momentum takes the following
form. 
\begin{equation}
P = P_\textrm{osc} - kn - \sum_{i=1}^4 m_i u_i
\end{equation}
We first insert
\begin{equation}
 \delta_{P,0} = \int_0^1 dt e^{2\pi i t P}
\end{equation}
into the trace in \eqref{thermalPF} in order to pick up states
satisfying the level matching condition, and then sum over all
states of Hilbert space. The partition function becomes 
\begin{align}
 F(\beta) = -\sum_{l=1}^\infty \sum_{k=1}^\infty 
\sum_{n\in \textbf{Z}} 
\sum_{w^A \in \textbf{Z}^8}
  \sum_{0\leq m_i < N_ik} \frac{1}{\beta l} \textrm{tr} &
  \Biggl[ (-1)^{(l+1)\textbf{F}} \int_0^1 dt e^{2\pi i t 
 (P_\textrm{osc} - m_iu_i - kn)}\nonumber\\
 & \times e^{\frac{\beta l}{\sqrt{2}} 
 \left( \frac{R^-}{\alpha' k}(H_\textrm{osc} 
 +\frac{|R_{\{w^A\}}|^2}{2\alpha'} )
 + \frac{k}{R^-}\right)}\Biggr].
\label{eq73}
\end{align}
We decomposed the trace into the zero-mode part and
the oscillator-mode part, and explicitly represent the summation over the zero-modes
by $\sum$. So, the trace in (\ref{eq73}) means the summation only over the oscillator
modes. By computing the level matching part and the summation over $n$,
we obtain
\begin{equation}
\sum_{n\in \textbf{Z}} \int_0^1 dt e^{2\pi i t P} 
= \sum_{q\in \textbf{Z}_k} e^{2\pi i \frac{q}{k} (P_\textrm{osc} 
-  \sum_i m_i u_i)}.
\end{equation}
The factor $e^{-2\pi i \frac{q}{k} \sum m_i u_i}$ together with the
summation over $m_i$'s becomes
\begin{equation}
 \sum_{m_i \in (\textbf{Z}_{k N_i})^4} e^{-2\pi i \frac{q}{k}
  \sum_i m_i u_i} = \prod_i (k N_i \sum_{p_i \in \textbf{Z}} 
  \delta_{qu_i,kp_i}).
\end{equation}
The Kronecker's delta from the $i$-th direction
$\sum_{p_i} \delta_{qu_i,kp_i}$ restricts values of $u_i$ to
$\frac{k}{d}\textbf{Z}$, where $d$ is $\textrm{gcd}(k,q)$.
This is equivalent to the restriction of $w^A$ to $(k/d) \textbf{Z}$.

Finally, we obtain 
\begin{align}
F(\beta) = -
      \sum_{k=1}^\infty 
      \sum_{l=1}^\infty
\sum_{q=0}^{k-1}
\sum_{w^A \in [(k/d){\bf Z}]^8}
\frac{{\cal N}_{LLL}}{\beta l k} 
e^{-\frac{\beta^2 l^2}{4\pi \alpha' \tau_2}\left( 1+
\frac{{R^-}^2 R^2_{\{w^A\}}}{2{\alpha'}^2 k^2}\right)} 
\times (\textrm{oscillator part}),
\end{align}
where ${\cal N}_{LLL} = k^4 \prod_i N_i$.
Namely, the result for ${\bf S}^1$ compactification \eqref{S1TPF} 
is generalized to the ${\bf T}^8$ compactification by
replacing the summation over one winding number $w$
by the summation over the eight winding numbers $w^A$
and ${\cal N}$ by ${\cal N}_{LLL}$
The oscillator part is not affected by the compactification
and is written by massive theta function
according to the oscillator spectrum of the ${\bf T}^8$ case\cite{suga2}.
While we chose the basis $\{\vec{\tilde{b}}_A\}$ depending on
the winding direction during the
calculation, the final expression of the TPF does
not depend on this choice.

\section*{Acknowledgements}
We would like to thank T.~Kawano, Y.~Matsuo, Y.~Sugawara and H.~Takayanagi for helpful discussions.
Research of Y.I. is supported in part by
Grant-in-Aid for the Encouragement of Young Scientists (\#15740140) from the Japan Ministry of Education, Culture, Sports,
Science and Technology.

\end{document}